\documentclass{PoS}

\newcommand{\swift}{{\it Swift}}
\newcommand{\ax}{$\alpha_{\rm X}$}
\newcommand{\aox}{$\alpha_{\rm ox}$}
\newcommand{\auv}{$\alpha_{\rm UV}$}
\newcommand{\lledd}{$L/L_{\rm Edd}$}

\title{Statistical Analysis of an AGN sample with Simultaneous UV and X-ray
Observations with Swift}

\ShortTitle{Statistical Analysis of AGN}

\author{\speaker{Dirk Grupe}\\
        Pennsylvania State University, 525 Davey Lab, University Park, PA 16802,
	USA \\
        E-mail: \email{grupe@astro.psu.edu}}

\abstract{I report on the statistical analysis of a sample of about 100 AGN with
simultaneous UV and X-ray observations with \swift.
 I found clear correlations
between the X-ray spectral slope \ax, the UV slope \auv, and the
optical-to-x-ray spectral slope \aox with the Eddington ratio \lledd. I
also report on the bolometric corrections for $L_{\rm 0.2-2.0 keV}$
 and $L_{5100\AA}$. A
major aspect of the statistical analysis is multi-variate analysis
statistical tools such as the Principal Component Analysis (PCA) and cluster
analysis. This analysis shows that the main driver of the AGN properties in this
sample is the Eddington ratio \lledd. Although separating Seyfert 1s into NLS1s
and BLS1s is a good classification, with the 2000 km s$^{-1}$
cutoff line it is still
arbitrary. A better classification scheme may be to separate AGN into low and
high \lledd\ AGN as suggested from the cluster analysis.
}

\FullConference{Narrow-Line Seyfert 1 Galaxies and their place in the Universe - NLS1,\\
		April 04-06, 2011\\
		Milan Italy}

\begin{document}

\section{Introduction}

After the milestone paper by \cite{oster85} Narrow Line Seyfert 1 galaxies
(NLS1s) have become an important part in AGN research. Their extreme properties
place them at one end of the AGN population, often known as the eigenvector 1
relation in AGN (e.g. \cite{boroson92, sulentic00, grupe99, grupe04b}). This
eigenvector 1 is typically associated with the Eddington ratio \lledd\ which
seem to drive the properties of AGN. \lledd\ is therefore a crucial parameter to
describe AGN. However, only for a small fraction of AGN it is possible to
directly determine \lledd. For the majority of AGN we rely on secondary methods
like a bolometric correction (e.g. \cite{elvis94, vasudevan08}) which calculate
the bolometric luminosity based on the luminosity at the specific wavelength.

In these proceedings I discuss the bolometric correction based on simultaneous
UV and X-ray observations obtained by \swift. This study was performed on a
sample of 92 AGN \cite{grupe10} which will also show how spectral slope can be
used as an estimator of \lledd. When \cite{oster85} and \cite{goodrich89} described
NLS1s, they defined their definitions purely phenomenologically. After more than 25
years, is the definition of NLS1s just based on their FWHM(H$\beta$)$<$2000 km
s$^{-1}$ still valid, or do we have to extend this definition? Here multi-variate
analysis techniques 
come into play. With these techniques 
we can put the definition on a more solid statistical
basis.

\section{NLS1s in the Literature}

\begin{figure}[!h]
\includegraphics[width=0.6\textwidth]{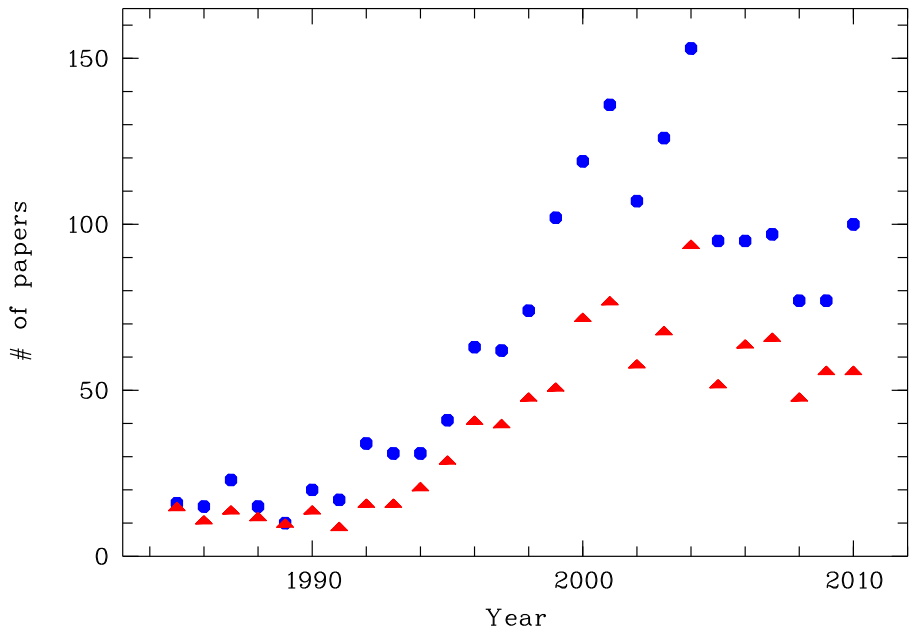}
\caption[ ] {\label{nls1_papers} 
Papers per year which have the name "NLS1" or "Narrow-Line Seyfert 1" in their
title or abstract. The red triangles display the refereed publications and the
blue circles show all publications.
} 
\end{figure}

When I gave a talk on NLS1 statistics at the 1999 Bad Honnef NLS1 meeting, I
started out showing how NLS1s had become more and more popular over the previous
decade and the number of publications with the word(s) "NLS1" or "Narrow-Line
Seyfert 1" in their title had grown exponentially \cite{grupe00}. Are NLS1s still
relevant in the literature today (March 2011) and more importantly, are NLS1s still
an active research field? 
Figure\,\ref{nls1_papers}
displays the numbers of papers per year have the name "NLS1" or "Narrow-Line Seyfert 1" in their
title or abstract.  The red triangles display the refereed publications and 
the blue circles show all publications. The number of NLS1 publications peaked
at around 2001/2002, but still is at a level of roughly 100 publications per
year including 50-60 refereed articles. Still, NLS1s are a very active field 
in the AGN community. 

Another support for this statement is that at the 1999 Bad Honnef NLS1 meeting
we had 60 participants. Of these only 6 are at this 2011 
meeting in Milan which has
more than 80 participants. So, in other words, NLS1s have attracted a lot of new
people in the field. As shown in the articles in these proceedings, the field of
NLS1s has really developed dramatically over the past ten years and we do have a
much better understanding now in 2011 what the place of NLS1s in the Universe is
than we had in 1999.

\section{Sample Selection}

Our AGN sample is selected from the ROSAT bright soft X-ray AGN sample by
\cite{grupe01, grupe04a} which contains a total of 110 AGN. There are several
advantages of this particular AGN sample. First, all AGN in the sample
are not only X-ray bright, but they are also bright in the Optical/UV which
allows us to obtain high quality data in a relatively short amount of observing
time. All of the AGN are intrinsically unabsorbed in X-rays and do not show high
reddening in the UV. By definition, all of these AGN have at least one ROSAT
observation which allows studies of their long-term variability. We also
obtained optical spectroscopy data for all AGN \cite{grupe04a} which enable us
to perform multi-variate statistical analysis between emission line and continuum 
properties. 

Of the 110 AGN, 92 had been observed by \swift\ by January 2010 and the results
of these observations were published in \cite{grupe10}. By Mid-2011 this sample
has been observed almost entirely. Half of the AGN in the sample are NLS1.
Almost all of the AGN have been observed in \swift's UV/Optical telescope (UVOT;
\cite{roming05}) in all 6 filters, which allows us to determine the Optical/UV
slope \auv\ and the majority of the AGN has been visited by \swift\ multiple
times, which  gives us a handle on the variability in these sources. Many of these
AGN show very strong variability in X-rays as well as in the UV \cite{grupe10},
which makes it necessary to observe the Optical/UV and X-ray bands
simultaneously when studying the spectral energy distributions of AGN.
The data analysis of the \swift\ observations of the AGN in the sample is
described in \cite{grupe10}.

\section{Swift}

\begin{figure}[!h]
\begin{center}
\includegraphics[width=0.6\textwidth]{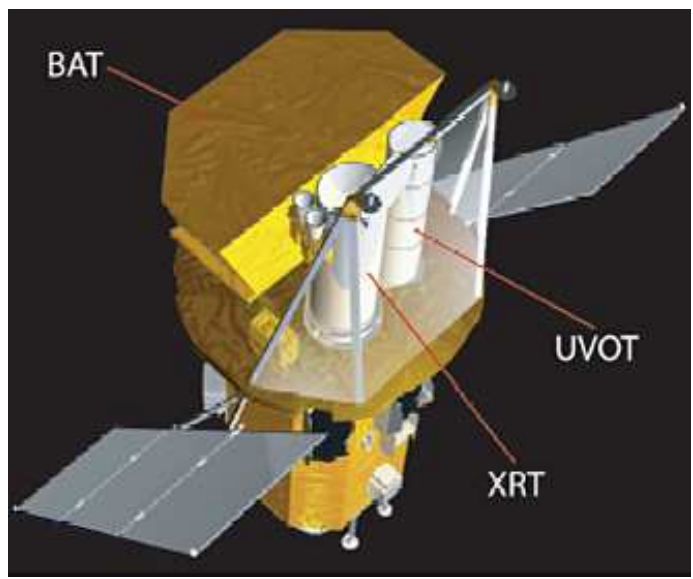}
\end{center}
\caption[ ] {\label{swift_tel} 
The \swift\ Gamma-Ray Burst Explorer Mission is equipped with the BAT, XRT and
UVOT. (Credit: NASA/GSFC)
} 
\end{figure}

The \swift\ Gamma-Ray Burst observatory was launched in November 2004
\cite{gehrels04}. While its prime objective was to detect and follow up on
Gamma-Ray bursts, \swift\ has turned into a multi-purpose observatory due to its
fast slewing and response capacity and its multi-wavelength coverage. 
As shown in Figure\,\ref{swift_tel}, 
\swift\ is
equipped with three telescopes, the Burst Alert Telescope (BAT;
\cite{barthelmy05}) which covers the 15-150 keV range, 
the X-ray telescope (XRT; \cite{burrows04}) covering the 0.3-10 keV energy band,
 and the UV/Optical Telescope (UVOT; \cite{roming05}) covering the 1800-6000\AA\
 wavelength range. The simultaneous X-ray and UV coverage makes \swift\ and
 ideal observatory for AGN.

\section{Simple correlation analysis}

The full simple statistical analysis of the \swift\ AGN sample can be found in
\cite{grupe10}. In these proceedings we only show a few of the correlations found
among the sample.

As mentioned in the introduction, one way to determine the bolometric luminosity
and therefore \lledd\ is through bolometric corrections. From our sample we
found the following bolometric corrections for the luminosity at 5100\AA\ and the
rest-frame 0.2-2.0 keV band:

\begin{equation}
\log L_{\rm bol}~=~(1.32\pm0.06)\times \log L_{\rm 5100}~-~(10.84\pm2.21).
\end{equation}

\begin{equation}
\log L_{\rm bol}~=~(1.23\pm0.06)\times \log L_{\rm X}~-~(7.36\pm2.01).
\end{equation}

These relations are displayed in Figure 17 in \cite{grupe10}.

\begin{figure*}[!h]
\includegraphics[width=0.6\textwidth]{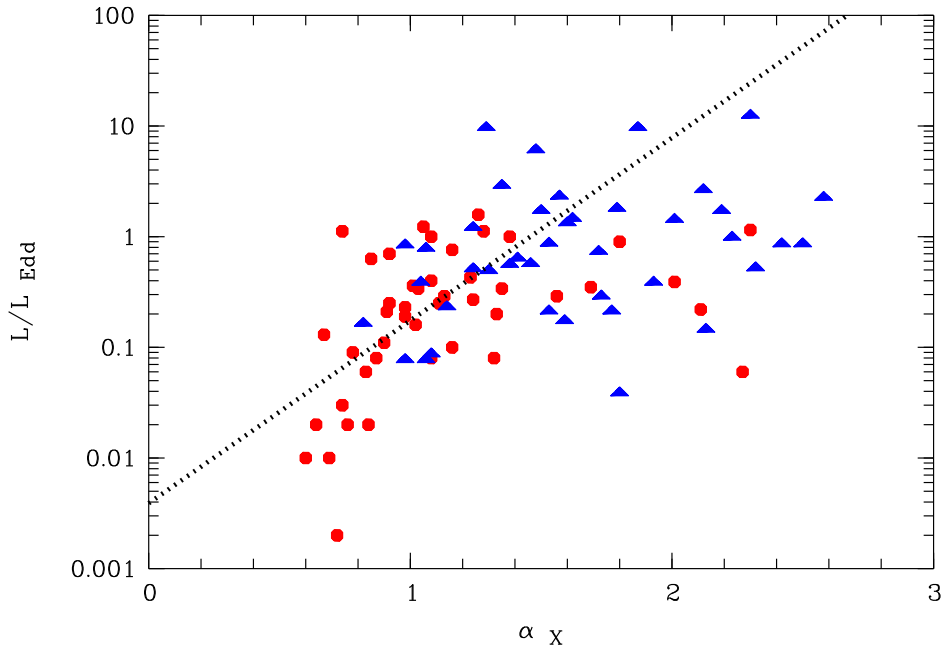}
\caption[ ] {\label{lledd_ax} 
Correlations between the Eddington ratio \lledd\ and the X-ray spectral slope
\ax. NLS1 are displayed as blue triangles and BLS1s as red squares.
} 
\end{figure*}

In particular we were interested in relations between \lledd\ and spectral slopes. 
Figure\,\ref{lledd_ax} displays the relation between
the X-ray spectral slope and \lledd. This is a strong correlation which is
actually dominated by the Broad Line Seyfert 1. NLS1s with the \lledd\ at unity
do not show a dependence of the X-ray spectral slope with \lledd. The relation
between \lledd\ and \ax\ found among the \swift\ AGN sample confirms earlier
results by \cite{grupe04a} using ROSAT results and \cite{shemmer08} using the
2-10 keV spectral slopes that high \lledd\ AGN have steeper X-ray continua than low
\lledd\ AGN.

\begin{figure*}[!h]
\includegraphics[width=0.6\textwidth]{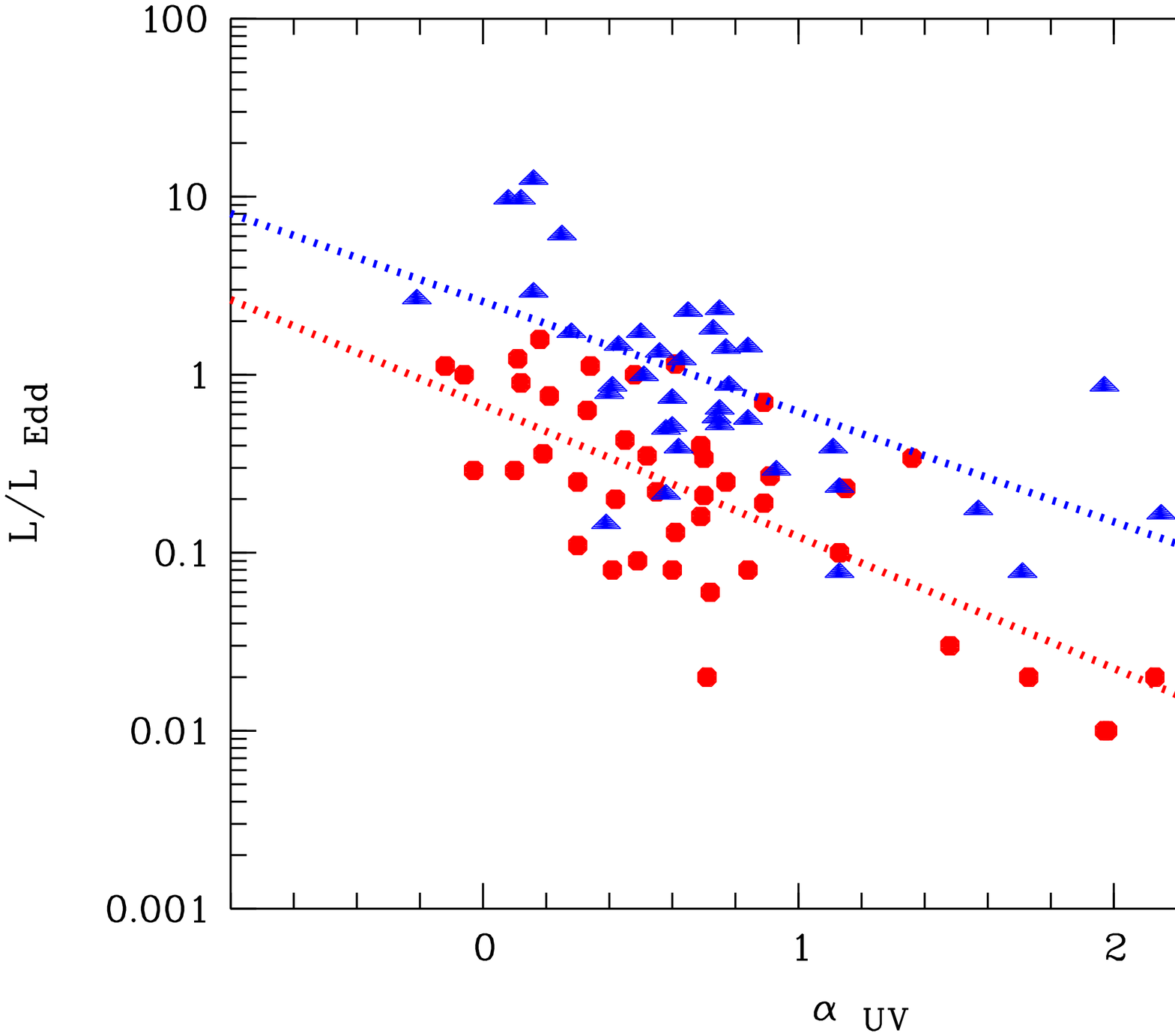}
\caption[ ] {\label{lledd_auv} 
Correlations between the Eddington ratio \lledd\ and 
 the Optical/UV spectral slope \auv. NLS1 are displayed as blue triangles and BLS1s as red squares.
} 
\end{figure*}

\begin{figure*}[!h]
\includegraphics[width=0.6\textwidth]{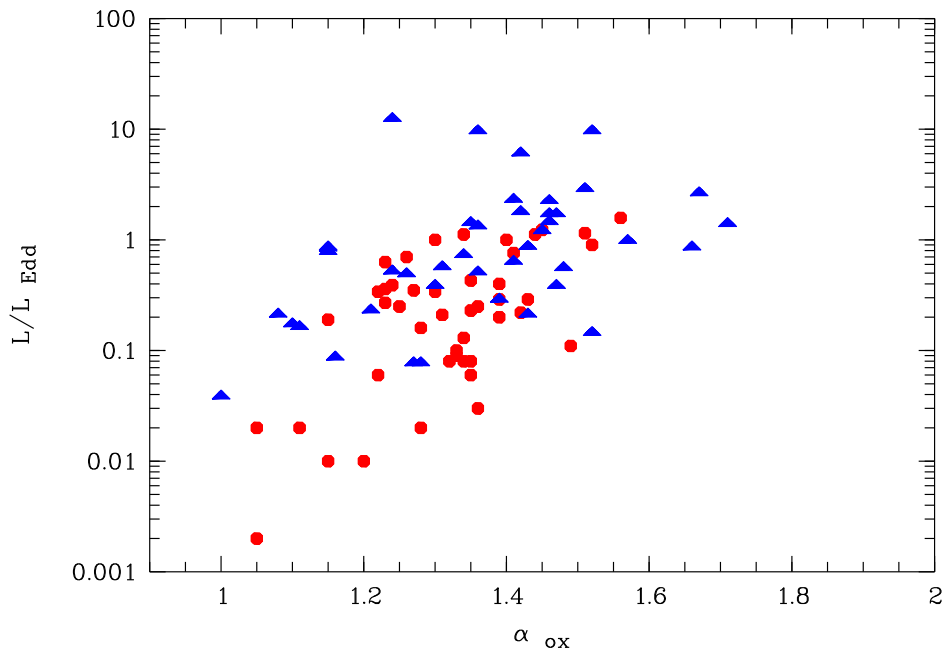}
\caption[ ] {\label{lledd_aox} 
Correlations between the Eddington ratio \lledd\ and 
optical to X-ray spectral slope
\aox. NLS1 are displayed as blue triangles and BLS1s as red squares.
} 
\end{figure*}

Figure\,\ref{lledd_auv} shows the anti-correlation
between \lledd\ and the Optical/UV spectral slope \auv. AGN with higher \lledd\
display bluer optical/UV spectra. Note that there is an offset between NLS1s and
BLS1s with NLS1s showing higher Eddington ratio for a given UV spectral slope
than BLS1s \cite{grupe10}. We also notices a strong correlation between \lledd\
and the optical to X-ray spectral slope
\aox\footnote{The X-ray loudness is defined by \cite{tananbaum79} as
  \aox=--0.384 $\log(f_{\rm 2keV}/f_{2500\AA}$).} as shown in
  Figure\,\ref{lledd_aox}. AGN with higher \lledd\ appear to be X-ray weaker
  \cite{grupe10}.

\section{Multi-variate Analysis}
Instead of only looking for linear relations between measured parameters in a
sample, multi-variate analysis examines the full data set in multiple
dimensions. Multi-variate analysis looks into the n-dimensional parameter space and
is searching for underlying properties/components that drives the observed
parameters. It also searches for groups/clusters in this parameter space and by
inter-weaving the different methods is able to find new relations and classes
within the data set. 

\subsection{Statistical software R}
All of the multi-component statistical analysis methods 
shown here  used the statistical package {\bf R} (e.g. \cite{crawley07,
rteam09}).
 {\bf R} is an object-oriented language for statistical data analysis. 
 It is an extremely
powerful package and 
unlike commercial statistical software packages, 
{\bf R} is a free package that is continuously maintained and developed 
\footnote{http://cran.r-project.org}. {\bf R} provides a huge library of statistical
analysis packages that can be used for astronomical purposes.

\subsection{Principle Component Analysis}

The Principal Component Analysis (PCA)
is a relatively old statistical tool, already developed more than
100 years ago (Pearson 1901). The idea of the PCA is to reduce the number of
parameters in a sample to a small number of significant parameters that capture
most of the variance in the data. For example when we look at the parameters that 
can be measured from stars
they can basically be described by their mass and  metallicity. 
In a mathematical sense, the PCA  searches for the
eigenvalues and eigenvectors in a correlation coefficient matrix:

\begin{equation}
A\times x = \lambda \times x
\end{equation}

where $A$ is the correlation coefficient matrix, $\lambda$ an eigenvalue  and
$x$ an eigenvector. In a geometrical sense, 
the PCA performs an axis transformation  for the cloud of data
points in this n-dimensional space  (n input parameters/properties). 
The first principal component is the eigenvector which
direction is along the longest elongation of this cloud. The second eigenvector
is the vector perpendicular this this one in the direction of the second
largest elongation of the data cloud, and so forth. 
A good description for the application of a PCA in astronomy
can be found in \cite{francis99, boroson92}. Always keep in mind that a PCA is a
purely mathematical method that can but does not necessarily result in physically
meaningful parameters.

\begin{figure}[!h]
\begin{center}
\includegraphics[width=0.6\textwidth]{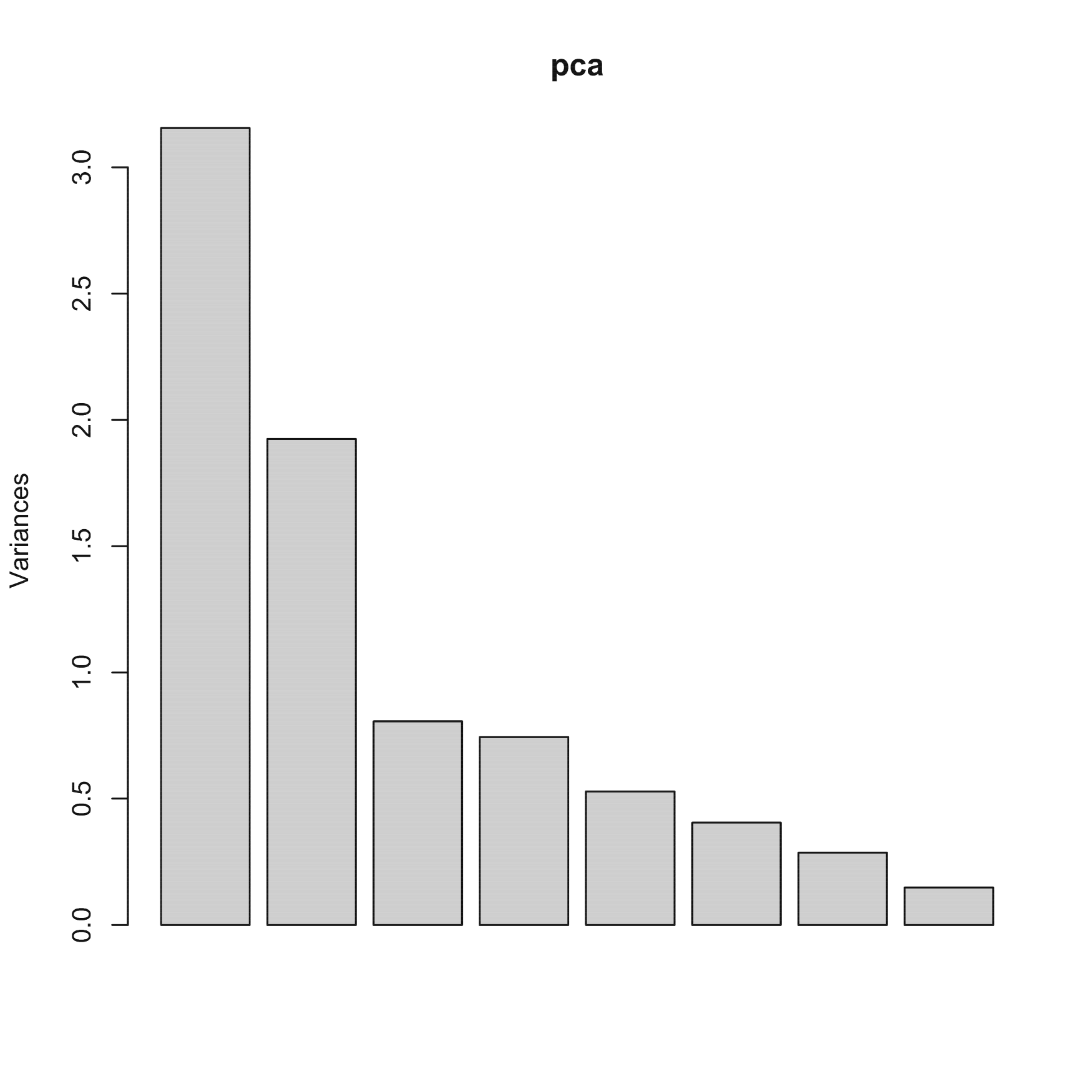}
\end{center}
\caption[ ] {\label{pca_plot} 
Strength of the eigenvectors from the Principal Component Analysis of our AGN
sample. The columns represent each eigenvector starting with eigenvector 1 at the
left.
} 
\end{figure}

\begin{figure}[!h]
\includegraphics[width=0.6\textwidth]{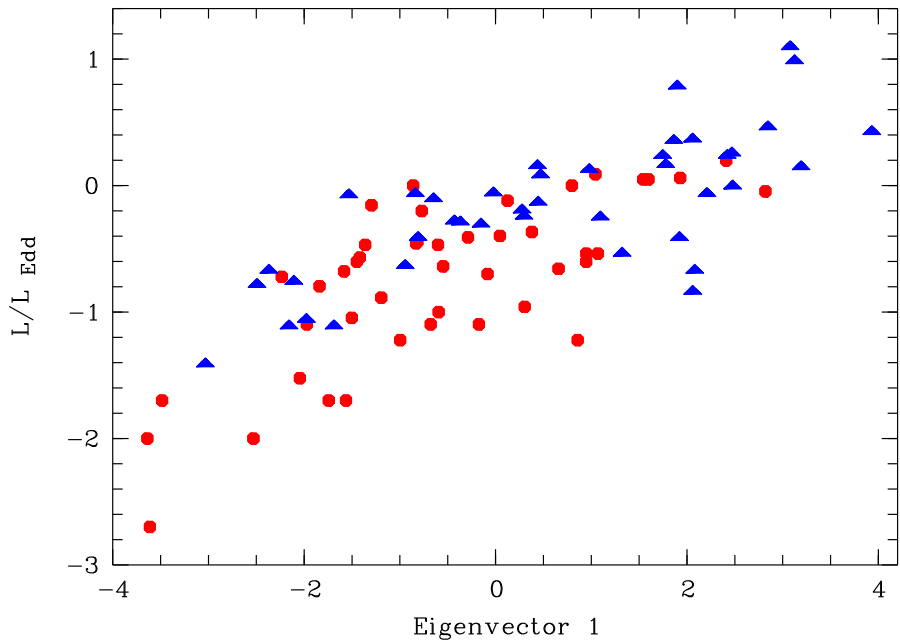}
\caption[ ] {\label{ev1_lledd} 
Eigenvector 1 vs. \lledd, again NLS1s are displayed as blue triangles and BLS1s as
red squares.
} 
\end{figure}

When applying the PCA to the AGN sample by using \ax, \auv, \aox, FWHM(H$\beta$),
\linebreak
FWHM([OIII]), [OIII]/H$\beta$,
FeII/H$\beta$, and $L_{\rm 0.2-2.0 keV}$ as input parameters we found
that the first two eigenvectors already account for 64\% of the variance in the
sample. Figure\,\ref{pca_plot} displays the strengths of the eigenvectors. A 
stronger eigenvector 1 results is steeper X-ray spectra, bluer optical spectra,
steeper \aox. We also found the well-known eigenvector 1  [OIII] and
FeII Boroson \& Green \cite{boroson92} relation. These relations suggests that
eigenvector 1 is associated with \lledd. Figure\,\ref{ev1_lledd} displays the
relation between eigenvector 1 and \lledd\ and indeed it shows there is a very strong
correlation between these two properties. Looking at eigenvector 2,
 we interpret this
eigenvector with the mass of the central black hole.

\subsection{Cluster Analysis}

\begin{figure}[!h]
\includegraphics[width=0.8\textwidth]{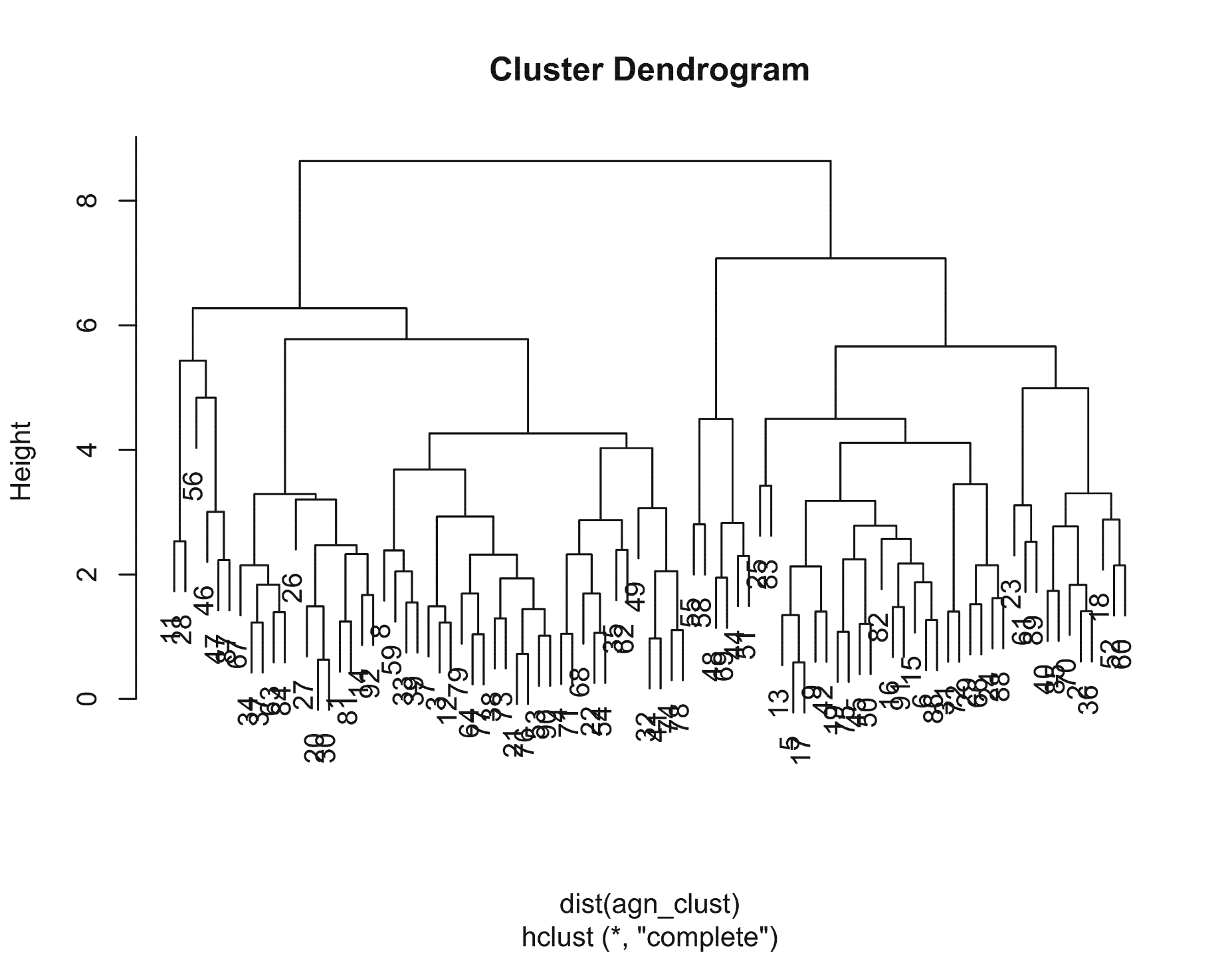}
\caption[ ] {\label{cluster_plot} 
Dendrogram of the hierarchical cluster analysis of the AGN sample. Clearly the AGN
sample can be divided into two groups. 
} 
\end{figure}

The other multi-component technique
is cluster analysis.
Cluster analysis is a purely statistical/mathematical tool that allows an 
unbiased
search for groups in multi-dimensional data sets.
In its simplest
form, the cluster analysis will search for the nearest neighbor 
in the parameter
space. It will then group these neighbors with other groups of neighbors and so
on. This is also called hierarchical clustering (e.g. \cite{duda01}).
At the end we will get a so called dendrogram such as displays for our
AGN sample in Figure\,\ref{cluster_plot}. 
Clustering is a powerful tool for
data mining. It can and will find groups and correlations that are otherwise
missed with simple analysis and classifications tools. A 
consistency-check of the
results of the cluster analysis can be done again through the PCA. 
The groups found in the cluster analysis should also be found as groups in the
eigenvector 1 - eigenvector 2 diagram from a PCA.

\begin{figure}[!h]
\includegraphics[width=0.6\textwidth]{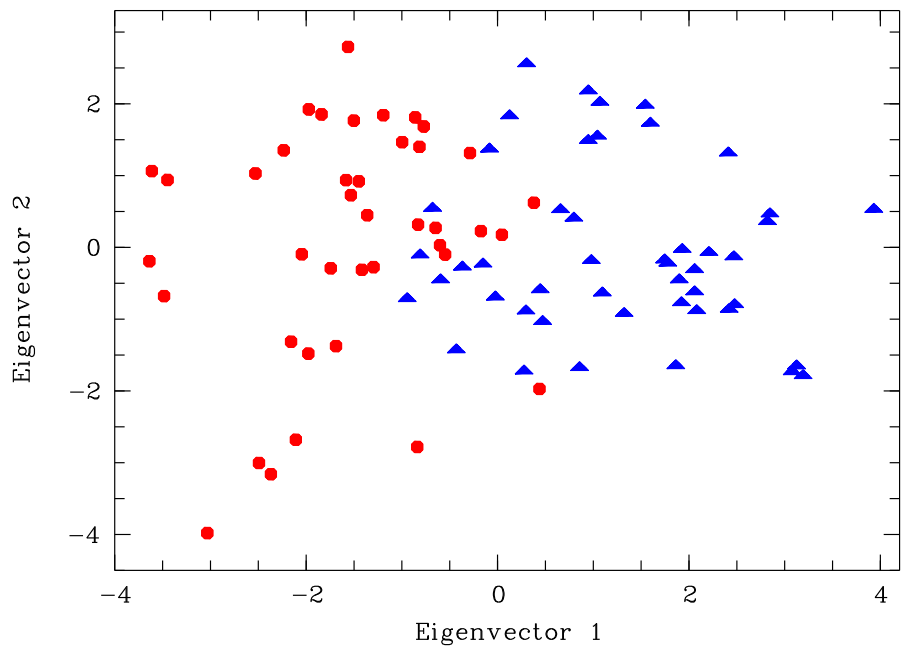}
\caption[ ] {\label{ev1_ev2_cluster} 
Eigenvector 1 vs. eigenvector 2 from the PCA using the two main groups found by the
cluster analysis. Group 1 is displayed as blue triangles and group 2 as red squares.
} 
\end{figure}

Figure\,\ref{cluster_plot} displays the dendrogram for our AGN sample using the
sample input parameters as used for the PCA. Clearly the AGN can be separated into
two classes. By itself, cluster analysis is a purely mathematical method and these
groups may not mean anything. However, let's have a look what the properties of the
two groups are. Group 1 shows steeper X-ray spectra, bluer optical/UV continua and
steeper optical to X-ray spectral slopes \aox. It also contains mostly objects with
narrower H$\beta$ lines and higher \lledd\ Eddington ratios. The AGN in group 1 are
those with the strongest FeII and weakest [OIII] emission. Does this mean that group
1 contains mostly NLS1s and group 2 BLS1s? Let's have at first a look at the
eigenvector 1 and eigenvector 2 diagram as we found it from the PCA. This graph is
shown in Figure\,\ref{ev1_ev2_cluster}. The members of group 1 in this plot are
shown as (blue) triangles. These are the objects with high eigenvector 1s and as we
have seen before in the PCA, these are the objects with high \lledd. While 62\% of
the members in group 1 are NLS1s, only 20\% of the members of group 2 are NLS1s,
which is more the typical fraction of NLS1s found among AGN.

Finally we can ask if the cutoff line between NLS1s and BLS1s at 2000 km s$^{-1}$ 
is still a valid definition. The answer is yes and no. No, because the 2000 km s$^{-1}$
definition is purely arbitrary and there are many sources with NLS1-like
properties, but which do have broader FWHM(H$\beta$) than the 2000 km s$^{-1}$
definition. Yes, because a majority of NLS1s fall into our group 1. All in all,
NLS1 is still a valid class/definition, but we should be more flexible about the
cutoff line. We should generally speaking talk more about high \lledd\ AGN because
this definition is based on a more statistical and physical definition.

\section{Discussion and Conclusions}

Although NLS1 research is now more than 25 years old, it is still an active field
in AGN research with still creates a lot of interest. It remains to be a powerful
research area which generates more than 50 refereed publications each year.
To make it clear, 
{\bf NLS1s are not an historical accident!} NLS1s are extreme AGN, but they are
needed for our interpretation of the AGN phenomenon. 

Extending our statistical study into n-dimensional parameter space by using
multi-variate tools like the PCA and cluster analysis shows that AGN separate into
high and low \lledd\ objects. The high \lledd\ sample is mostly represented by
NLS1s. Although the 2000 km s$^{-1}$ cutoff line between NSL1s and BLS1s is
somewhat arbitrary, NLS1 still remains to be a valid AGN class.

\acknowledgments
Swift is supported at Penn State by NASA contract NAS5-00136. 
This research has been supported by NASA contracts NNX08AT25G, 
NNX09AP50G and NNX07AH67G..

\end{document}